\documentclass[%
 reprint,
 amsmath,amssymb,
 aps,
]{revtex4-1}

\usepackage{graphicx}
\usepackage{dcolumn}
\usepackage{bm}
\usepackage{amsmath}
\usepackage{xcolor}

\begin{document}

\preprint{APS/123-QED}

\title{Observation of transition from semiconducting to metallic ground state in high-quality single crystalline FeSi}
\author{Y. Fang$^{1,2}$}
\author{S. Ran$^{3,4}$}
\author{W. Xie$^{5}$}
\author{M. B. Maple$^{1,2,3}$}
\email[Corresponding Author: ]{mbmaple@ucsd.edu}

\affiliation{$^1$Materials Science and Engineering Program, University of California, San Diego, La Jolla, California 92093, USA}
\affiliation{$^2$Center for Advanced Nanoscience, University of California, San Diego, La Jolla, California 92093, USA}

\affiliation{$^3$Department of Physics, University of California, San Diego, La Jolla, California 92093, USA}
\affiliation{$^4$Department of Materials Science and Engineering, University of Maryland, Collage Park, 20742, USA}
\affiliation{$^5$Department of Chemistry, Louisiana State University, Baton Rouge, LA, 70803, USA}
\date{\today}

\begin{abstract}
We report anomalous physical properties of single-crystalline FeSi over a wide temperature range 1.8-400 K. X-ray diffraction, specific heat, and magnetization measurements indicate that the FeSi crystals synthesized in this study are of high quality with a very low concentration of magnetic impurities ($\sim$0.01$\%$). The electrical resistivity $\rho$($T$) can be described by activated behavior with an energy gap $\Delta$ = 57 meV between 67 K and 150 K.  At temperatures below 67 K, $\rho$($T$) is significantly lower than an extrapolation of the activated behavior, and the Hall coefficient and magneto-resistivity undergo a sign change in this region. At $\sim$19 K, a transition from semiconducting to metallic-like behavior is observed with deceasing temperature. Whereas the transition temperature is very robust in a magnetic field, the magnitude of the resistivity below $\sim$30 K is very sensitive to magnetic field.  There is no indication of a bulk phase transition or onset of magnetic order in the vicinity of either 67 K or 19 K from specific heat and magnetic susceptibility measurements. These measurements provide evidence for a conducting surface state in FeSi at low temperatures.
\end{abstract}

\pacs{Valid PACS appear here}

\maketitle

\section{Introduction}

The transition metal silicides FeSi, MnSi, CoSi, and CrSi, have the B20 crystal structure, which is the only group in the cubic system without an inversion center. These compounds exhibit rich physical phenomena that are of great interest for fundamental understanding and potential applications. For example, the $d$-electron compound FeSi shows a remarkable similarity to $f$-electron Kondo insulators, and the electrical resistivity $\rho$($T$) evolves continuously with decreasing temperature from metallic behavior (d$\rho$/d$T$ $>$ 0) to strongly correlated semiconducting behavior (d$\rho$/d$T$ $<$ 0)~\cite{jaccarino1967paramagnetic,schlesinger1993h,sales1994magnetic,　mandrus1995thermodynamics,paschen1997low}. A considerable amount of theoretical effort~\cite{takahashi1979theory,aeppli1992comments,fisk1995kondo,varma1994aspects,fu1995model,anisimov1996singlet} has been expended to explain the strong temperature dependence of the magnetic susceptibility $\chi$($T$) of FeSi which reaches a maximum value at around 500 K~\cite{jaccarino1967paramagnetic} that is not related to magnetic order~\cite{wertheim1965unusual,kohgi1981neutron}.




The ground state of FeSi is considered to be non-magnetic; however, experimental investigations of FeSi at low temperature reveal features that are sample dependent and are not well understood~\cite{paschen1997low,jarlborg1995low,sun2014resonant}. The published results are consistent in terms of the small semiconducting energy gap of 50-60 meV in the temperature range of 70-170 K. However, further decrease of the temperature results in either saturation steps~\cite{paschen1997low}, a hump (shoulder) at 70 K~\cite{sun2014resonant,petrova2010elastic} or $\sim$ 35 K~\cite{mihalik1996magnetic}, a moderate increase of $\rho$ with decreasing temperature below 40 K~\cite{bocellip1995evolution} or 50 K~\cite{schlesinger1993unconventional}, or a saturation below about 5 K~\cite{hunt1994low} of $\rho$($T$). Besides, the values of $\rho$ below 70 K reported by these references are also very different, indicating strong sample dependence of the electrical transport behavior. It has been well established in experiments that the electrical properties of semiconductors can be very sensitive to external dopants.~\cite{sales2011thermoelectric,dietl2010a,fang2015chemical}  To investigate the intrinsic physical properties of FeSi, we prepared high quality single-crystal samples of FeSi and performed various physical property measurements over a wide temperature range of 1.8-400 K. Anomalous electrical transport behavior associated with a change in primary charge carriers and negative magneto-resistivity at low temperatures were observed in all of the samples. We also report metallic conducting behavior of FeSi single crystals below $\sim$19 K, yielding evidence for a conducting surface state, consistent with specific heat, magneto-resistivity, and magnetization measurements.

 
\section{Experimental details}
Single-crystalline samples of FeSi were grown in high-temperature Sn flux with Fe: Si molar ratio of 1: 1.
The quality of the FeSi samples was assessed by means of single crystal X-ray diffraction at room temperature. A Bruker Apex II X-ray diffractometer with Mo K$_\alpha$$_1$ ($\lambda$ = 0.71073 \AA) radiation was used to measure the scattering intensity.  The crystal structure was refined with SHELXTL package~\cite{sheldrick2008short}. Electrical resistivity, magneto-resistivity, Hall effect, and specific heat measurements were performed in a Quantum Design Physical Property Measurement System (PPMS) DynaCool. The magnetization and magnetic susceptibility measurements were carried out in a Quantum Design Magnetic Property Measurement System (MPMS)~\cite{fang2015enhancement}.


\section{Results and Discussion}

\begin{figure}
\centering
\includegraphics[width=0.5\textwidth]{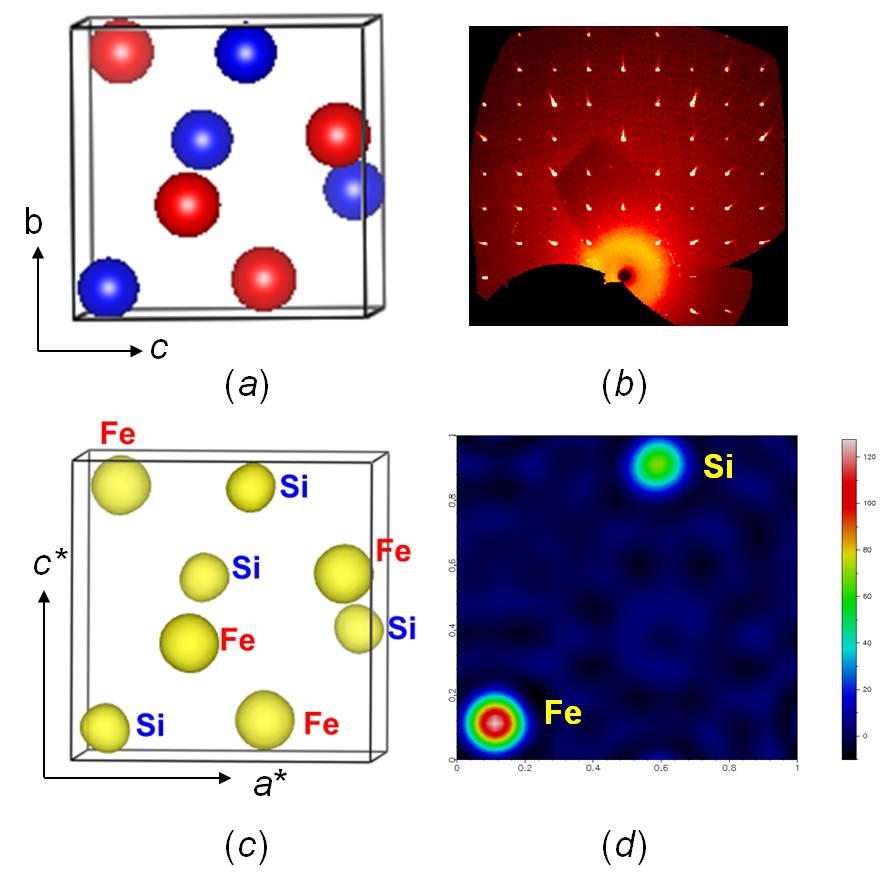}
\caption{(a) The crystal structure of FeSi (Red: Fe; Blue: Si). (b) Single crystal X-ray diffraction precession image of the ($h$0$l$) plane in the reciprocal lattice of FeSi at 300 K. All of the resolved spots correspond to the cubic chiral crystal structure $P$2$_1$3. (c) 3D Fourier map showing the electron density in $B$0-FeSi. (d) 2D Fourier map showing the electron density on Fe and Si along the $z$-axis.}
\end{figure}

The FeSi single crystals grow along the [111] direction in the Sn flux, resulting in bar-shaped samples. The results of single-crystalline X-ray diffraction on FeSi are shown in Fig. 1. Consistent with previous studies, stoichiometric FeSi crystallizes in the cubic chiral structure with space group $P$2$_1$3 ($B$20-type) and lattice parameter $a$ = 4.4860(5) \AA. No vacancies were observed according to the refinement. The resulting profile residual Rp is 1.79$\%$ with weighted profile residual Rwp 4.11$\%$. No electron density residual was detected, indicating the high quality of the FeSi crystals.

\begin{figure}
\centering
\includegraphics[width=0.5\textwidth]{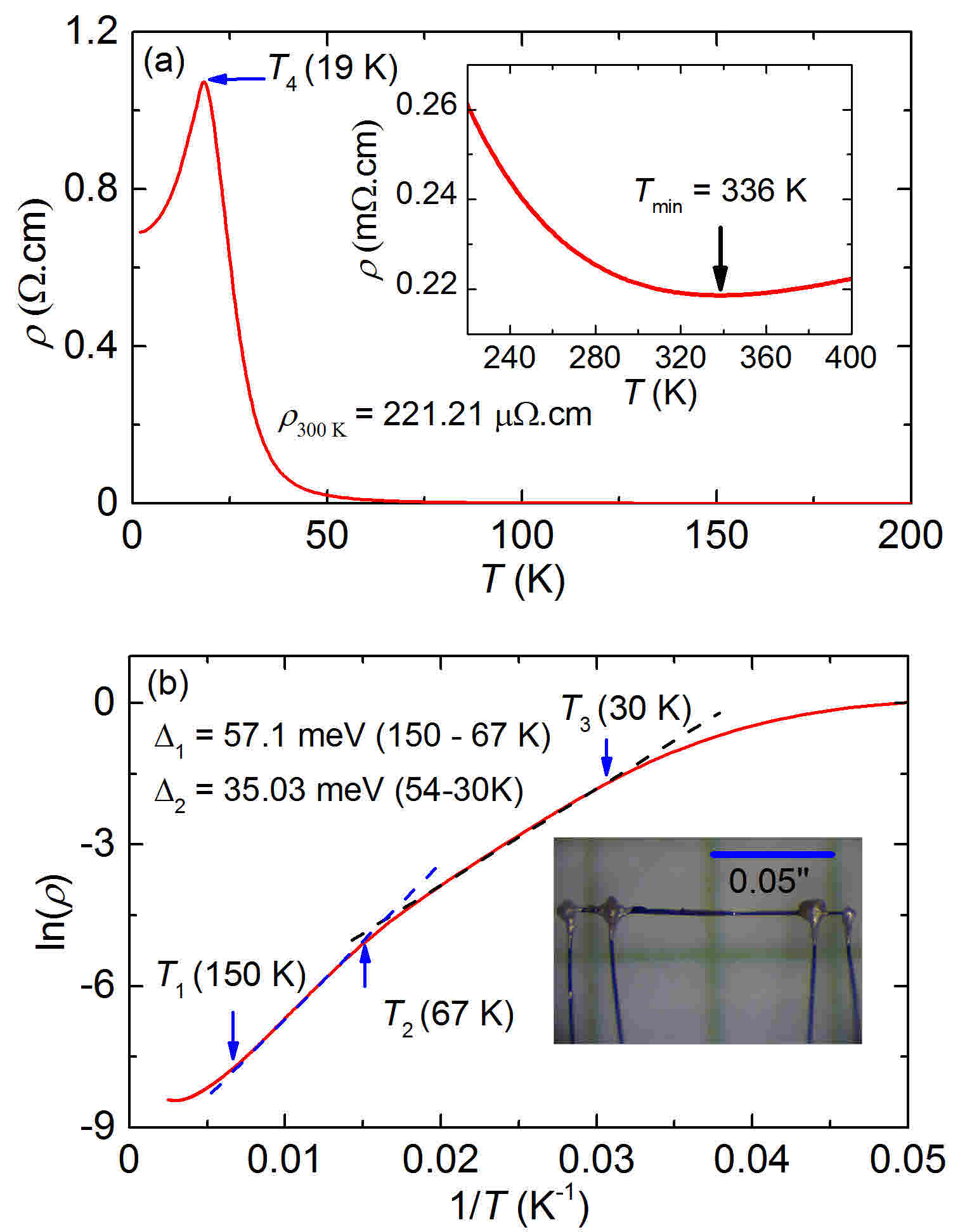}
\caption{(a) Electrical resistivity $\rho$ vs. temperature $T$ for FeSi with the current flowing along the [111] direction below 200 K. (b) ln($\rho$) vs. 1/$T$. The insets of (a) and (b) show the resistivity at high temperatures and a picture of the sample with the four-wire electrical lead configuration, respectively.}
\end{figure}

Because of their bar-shape and high quality, the FeSi single crystals are very suitable for electron transport measurements along the [111] direction. Upon cooling from 400 K, metallic-like behavior can be observed down to 336 K, below which the resistivity increases with decreasing temperature, resulting in a minimum in $\rho$($T$) at $T_{min}$ = 336 K   (see the inset of Fig. 2(a)). Similar features can also be found in other references with values of $T_{min}$ mostly in the range 150-300 K)~\cite{samuely1996gap,wolfe1965thermoelectric,buschinger1997transport,paschen1997low}. Decreasing the temperature further results in a gradual enhancement of semiconducting behavior down to 152 K, which has been reported to be related to the opening of a semiconducting energy gap ~\cite{ishizaka2005ultraviolet,arita2008angle,klein2008evidence}.

A plot of ln($\rho$) vs. (1/$T$) for FeSi shown in Fig. 2(b) is linear in the temperature range 152 K ($T_1$) to 67 K ($T_2$), consistent with standard activated behavior with an energy gap $\Delta$ = 57.1 meV; this value of $\Delta$ is comparable to previously reported gap values of 50-60 meV ~\cite{petrova2010elastic,schlesinger1993unconventional,jaccarino1967paramagnetic,sales1994magnetic}. From 54-30 K, where the relation ln($\rho$) vs. 1/$T$ is also linear, the value of $d$ln($\rho$)/d(1/$T$) corresponds to am energy gap of 35 meV. Below 30 K ($T_3$), the $\rho$($T$) curve cannot be described by a standard activation model. Further decrease of the temperature below 19 K ($T_4$) results in a decrease of $\rho$ with decreasing temperature as shown in Fig. 2(a). As the phenomena observed in $\rho$($T$) below $T_2$ are quiet different from the electrical transport behavior of FeSi reported in other references, we repeated the measurements on five different FeSi single crystals which yielded the same results. 

For a better understanding of the temperature dependence of the observed electrical transport behavior, especially the metallic conducting behavior, we performed specific heat $C_p$($T$) measurements on the samples down to 1.8 K, the results of which are shown in Fig. 3. The specific heat $C_p$($T$) can be reasonably well described by the sum of electronic and lattice contributions at low temperatures $C_p$ = $\gamma$$T$ + $\beta$$T^3$. No anomalies at $T_2$ = 67 K, $T_3$ = 30 K, and $T_4$ = 19 K can be observed, indicating the absence of any bulk phase transitions in FeSi at these temperatures. The estimated value of the Debye temperature $\theta$$_D$ of 457 K lies within the range of 377-515 K previously reported~\cite{paschen1997low,takahashi2000specific,marklund1974specific}. On the other hand, the electronic specific heat coefficient $\gamma$ is estimated to be 0.41 mJ.mol$^{-1}$.K$^{-2}$, which is only about 8-30$\%$ of previously reported values~\cite{paschen1997low}, suggesting that the samples studied in this work have a lower concentration of electron donor impurities, as the value of $\gamma$ is proportional to the density of electronic states at the Fermi level.  The metallic-like conduction below $T_4$ = 19 K exhibited by the FeSi samples in this work is dramatically different from the semiconducting behavior observed in other FeSi samples which have higher concentrations of charge carriers.  The seemingly contradictory phenomena suggest that the metallic conduction observed in this study is unlikely to be a bulk phenomenon, which is also supported by the absence of phase transition features in the $C_p$($T$) data.

\begin{figure}
\centering
\includegraphics[width=0.5\textwidth]{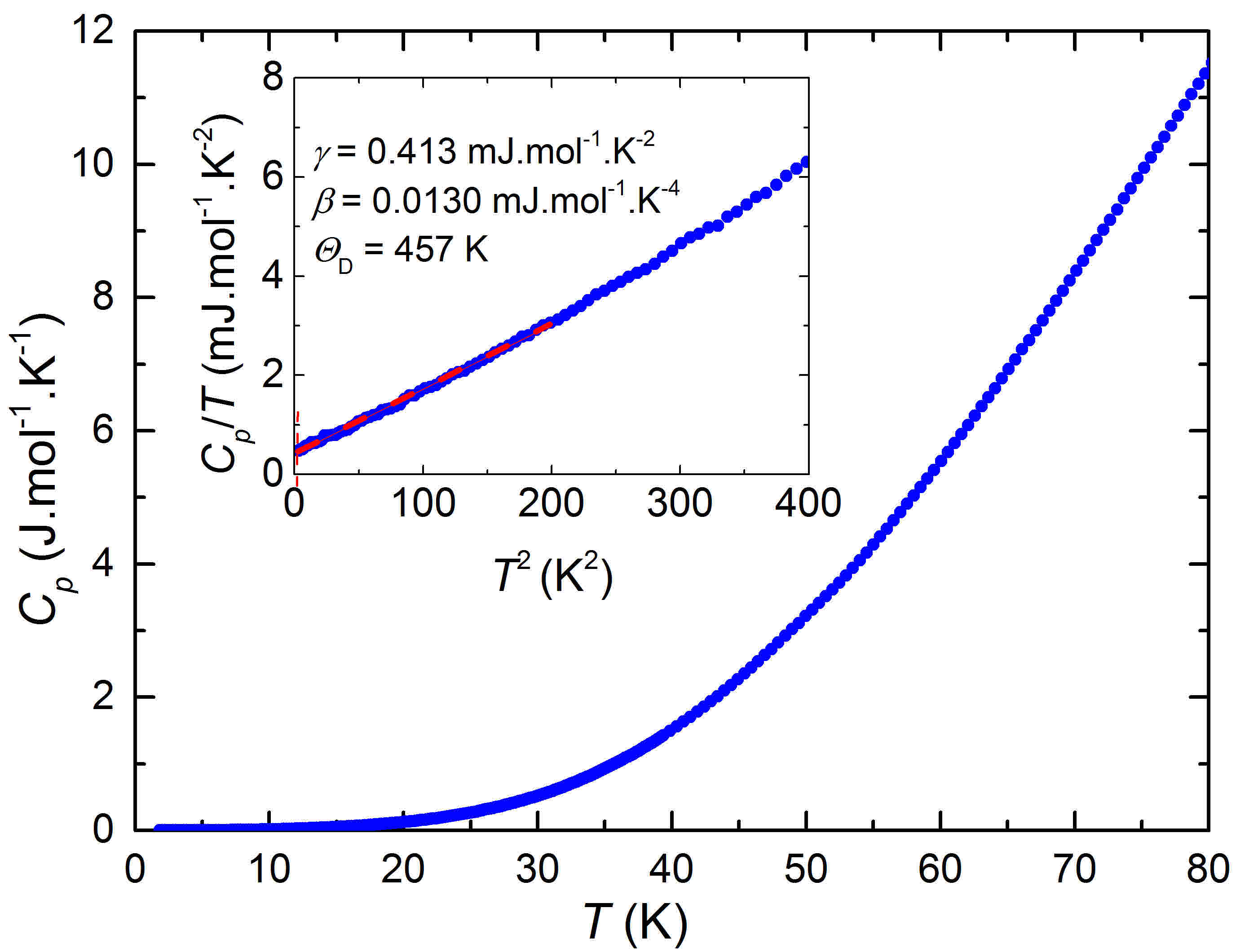}
\caption{Specific heat $C_p$($T$) of FeSi at low temperatures from 1.8 to 80 K. A plot of $C_p$/$T$ vs $T^2$ below 20 K is shown in the inset. The dashed line in the inset is a fit of the expression $C_p$/$T$ = $\gamma$
+$\beta$$T^2$ to the data with the values of $\gamma$, $\beta$, and $\theta$$_D$ given in the inset of the figure.}
\end{figure}

The temperature dependence of the magnetic susceptibility $\chi$($T$) for FeSi is shown in Fig. 4. Above 100 K, the value of susceptibility increases with increasing temperature, which is similar to the behavior of $\chi$($T$) for an antiferromagnet at temperatures below the N\'eel temperature. In the temperature range 20-100 K, $\chi$($T$) is very small $\sim$0.15 emu.mol$^{-1}$.T$^{-1}$, indicating a very weak response of FeSi to external magnetic field and a non-magnetic ground state for FeSi. Below 15 K, $\chi$($T$) of FeSi has a Curie-Weiss like upturn with decreasing temperature, which is believed to be associated with magnetic impurities~\cite{jaccarino1967paramagnetic,wertheim1965unusual}. In this study, however, the magnitude and the onset temperature of the $\chi$($T$) upturn is significantly smaller and lower, respectively, than previously reported values~\cite{koyama1999precise,mihalik1996magnetic,petrova2010elastic}, indicating lower magnetic impurity concentration for the present samples. The kink observed in the $M$($H$) curve at around 2 T indicated by the arrow in the inset of Fig. 4 is also consistent with the paramagnetic impurity scenario. Above 2 T, it seems that the magnetic field does not dramatically affect the magnetization of the samples, which also suggests the absence of magnetic order at low temperatures. The results of the magnetic measurements reveal that the samples are of high quality and the transitions observed around $T_2$ = 67 K and $T_4$ = 19 K in the $\rho$($T$) curve are not related to any bulk magnetic transitions.


In this study, the magnetization of FeSi can be well described by using the following Langevin functions: \begin{equation}
M = M_S[coth(\mu H/k_BT)-k_BT/\mu H]
\end{equation} in which $\mu$ is the magnetic moment of the impurity clusters and $M_S$ is the saturation magnetization. The corresponding fitting  of $M$($H$) at 3.5 K gives $M_s$ = 2.433$\times$10$^{20}$ $\mu$$_B$/mol and $\mu$ = 7.95 $\mu$$_B$. If we assume that the magnetic moment per impurity atom is 3$\mu_B$ as in pure iron, the concentration of impurity Fe atoms is only about 130 ppm per Fe atom, which is significantly lower than the impurity concentration previously estimated for single crystal specimens of FeSi~\cite{koyama1999precise,wertheim1965unusual}. The fitting results also show that, on average, there is only about 2-3 magnetic impurity atoms in each cluster, indicating atomic size magnetic clusters. 


\begin{figure}
\centering
\includegraphics[width=0.5\textwidth]{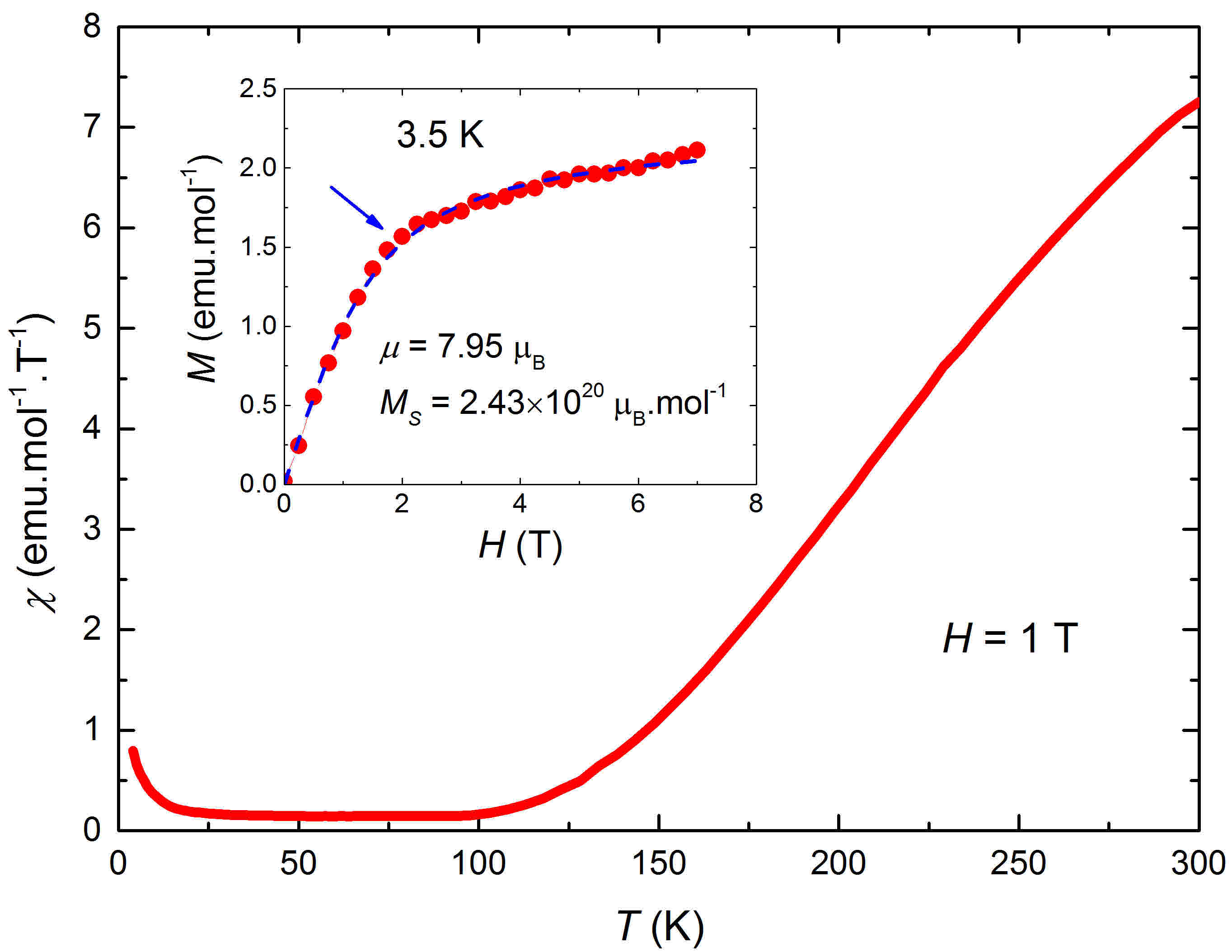}
\caption{Magnetic susceptibility $\chi$ vs. temperature $T$ for FeSi single crystals. The corresponding magnetization curve at 3.5 K is shown in the inset. The dashed curve is a fit of the Langevin function to the $M(H)$ data}
\end{figure}

Figure 5 shows $\rho$($T$) data for FeSi under external magnetic field. At high temperatures, the values of $\rho$ are almost independent of the applied magnetic field; however, a negative magneto-resistivity ($MR$), where $MR$ = ($\rho$$_{3 \rm T}$ - $\rho$$_{0 \rm T}$)/$\rho$$_{0 \rm T}$, can be observed around 70 K as indicated by the arrow in the inset of Fig. 5. that becomes very significant below 30 K, which is very close to the temperatures $T_2$ = 67 K and $T_3$ = 30 K, respectively. It should be mentioned that previous studies of the $MR$ are not consistent:  A negative $MR$ was reported by Paschen $et~al.$ below 30 K and attributed to quantum interference effects~\cite{paschen1997low}; however, a change of sign at around 70 K (close to $T_2$ = 67 K in this study) was reported later below which the $MR$ is positive~\cite{sun2014resonant}. In this study, the negative $MR$ reaches a minimum value at $T_4$ = 19 K. The peak in the absolute value of the $MR$ in this study is about 20$\%$, which is obviously higher than the peak in the absolute value of the $MR$ reported in Refs. \cite{paschen1997low} and \cite{sun2014resonant}, revealing the dependence of the $MR$ on sample quality. While $\rho$($T$) is very sensitive to the applied field at low temperatures, the value of $T_4$ seems independent of $H$, which provides additional evidence that the transition observed in $\rho$($T$) around $T_4$ is not related to magnetic order.

\begin{figure}[t]
\centering
\includegraphics[width=0.5\textwidth]{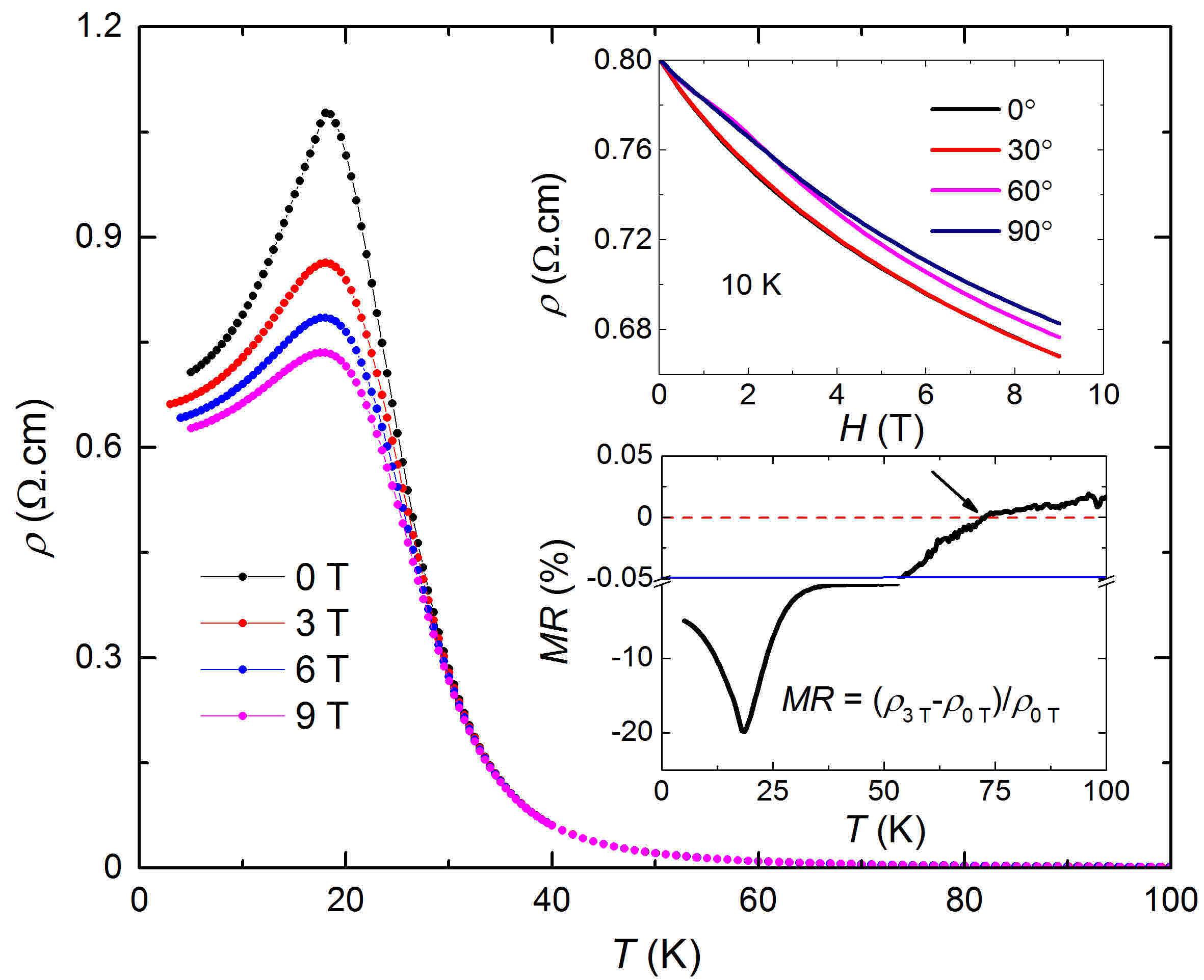}
\caption{Electrical resistivity $\rho$ vs. temperature $T$ in magnetic fields up to 9 T. The applied field is perpendicular to the current. Shown in the upper inset is $\rho$ vs. $H$ at several angles between the direction of the applied field and the current. Displayed in the lower inset is the temperature dependence of the magneto-resistance (MR). The definition of the MR is given in the lower inset.}
\end{figure}

\begin{figure}
\centering
 \includegraphics[width=0.5\textwidth]{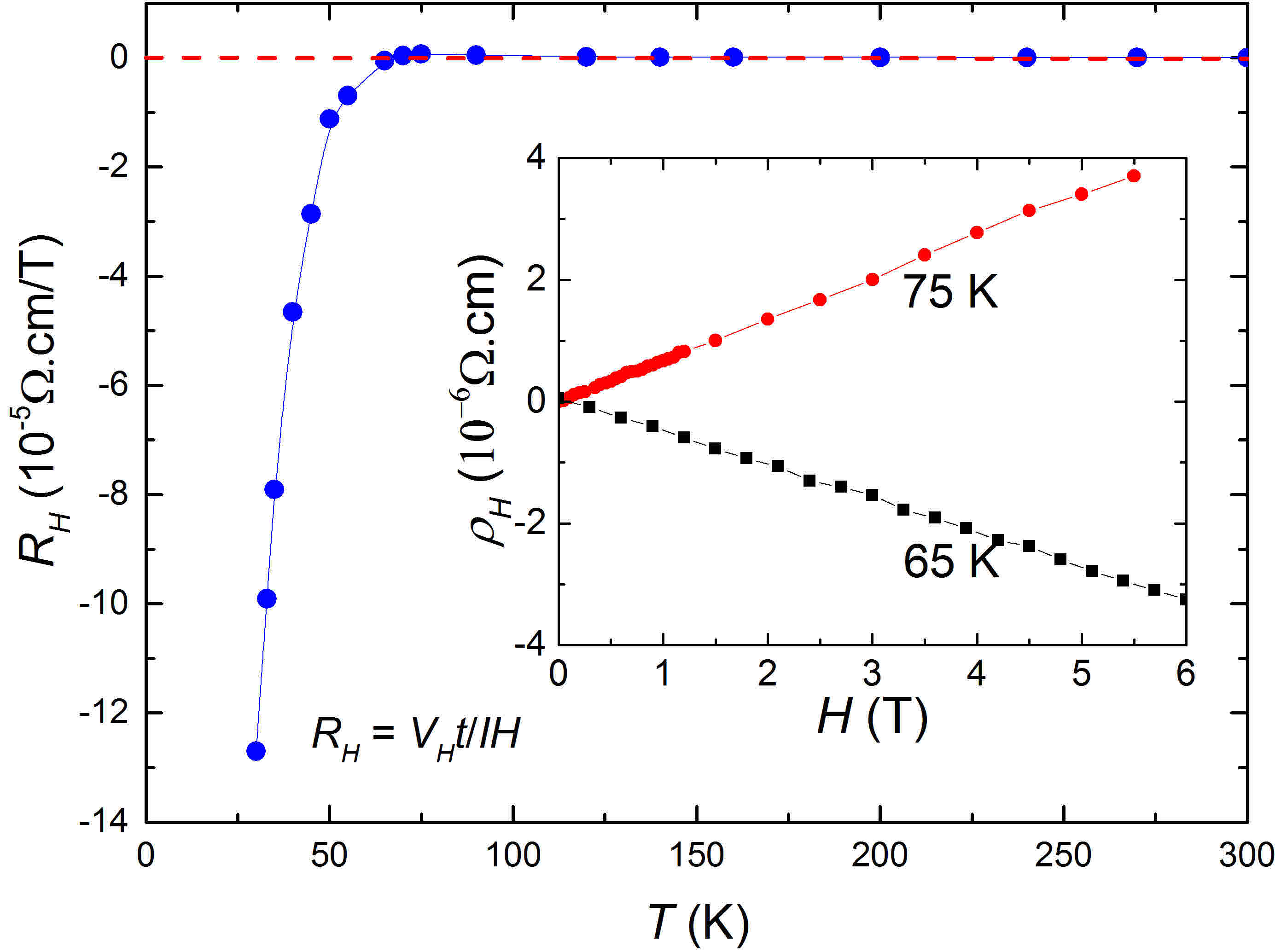}
\caption{Evolution of the Hall coefficient $R_H$ with temperature $T$. The Hall resistivity $\rho_H$ vs. $H$ at 65 and 75 K is shown in the inset.}
\end{figure}

The evolution of $\rho$ as a function of $H$ at several angles of $H$ with respect to the long axis of the FeSi crystal at 10 K is shown in the upper inset of Fig. 5. The value of $\rho$ is suppressed with increasing field, but the evolution of $\rho$($H$) deviates slightly from a linear relation. The angle between the directions of the applied field and the current ($\alpha$) has only a slight effect on the values of $\rho$($H$). However, the negative $MR$ is still very remarkable in the case that the applied field is parallel to the current (parallel to the surface of the sample), which is inconsistent with the behavior of topological insulators. This behavior can be qualitatively understood by considering both bulk and surface electron conduction for FeSi. The negative $MR$ can be observed at temperature $T_2$ which is far above $T_4$, suggesting that the negative $MR$ is a bulk phenomenon. If we assume that the response of surface resistivity to external field is positive due to the additional scattering of free electrons by the Lorentz force, increasing $\alpha$ will increase the effective applied field on the sample's surface and thus slightly enhance the $MR$.

The main results of the Hall effect measurements at temperatures down to 30 K are displayed in Fig. 6. Unlike the results of previous reports~\cite{paschen1997low}, linear relations of the Hall resistivity vs. applied external field can be seen up to 9 T over a wide temperature range above 30 K (see the inset of Fig. 6). At high temperatures, the Hall coefficient $R_H$ is positive and increases slightly with decreasing temperature, indicating that the dominant charge carriers are electrons (which is understandable as the mobility of electrons in intrinsic semiconductors are usually much higher than that of holes). However, a change of sign of $R_H$ is observed at $\sim$68 K, which is very close to $T_2$ = 67 K and to the temperature of the sign change of the $MR$. The phenomena observed in the $R_H$($T$), $MR$($T$), and $\rho$($T$) measurements are consistent with one another, indicating that there is a electronic phase transition at around $T_2$ = 67 K below which the resistivity is dominated by hole conduction and is sensitive to external field.

\section{Summary}

In summary, the following conclusions can be drawn from this study:

(1) The electron transport behavior of FeSi is highly sensitive to sample quality. High quality of the single crystal samples in this study is supported by X-ray diffraction, specific heat, and magnetization measurements. 

(2) In the temperature range 150-67 K, the semiconducting energy gap is 57 meV. Below 67 K, a  much smaller energy gap and a sign change of magneto-resistance and Hall coefficient are observed.

(3) Further decrease of the temperature results in a sharp reversible transition from a negative slope to a positive slope of $\rho$($T$) at 19 K. Corresponding magnetization and magneto-resistivity measurements suggest that there is no magnetic order associated with this transition. Furthermore, no feature can be observed in specific heat measurements. The results indicate that the metallic conduction behavior below $T_4$ is probably a surface phenomenon. 

(4) We should also emphasize that the intrinsic response of FeSi to an external magnetic field is almost zero below 100 K; however, a significantly large negative magneto-resistivity which reaches its maximum value at $T_4$ = 19 K is observed in the resistivity measurements. Considering the possibility that a trivial signal can not be observed in the MPMS,  the contradictory phenomena also suggests the existence of a special surface state that contributes additional electronic conductivity to the samples. 

(5) This study cannot provide a clear picture of the surface state of FeSi and the possibility that FeSi as a topological insulator cannot be ruled out. Further investigations of the electronic states on the surface are needed to explore this possibility. 

\section{Acknowledgements}
Helpful discussions with Zachary Fisk and James Allen are greatly appreciated. Materials synthesis and characterization at UCSD was supported by the US Department of Energy, Office of Basic Energy Sciences, Division of Materials Sciences and Engineering, under Grant No.~DEFG02-04-ER46105.  Low temperature measurements at UCSD were sponsored by the National Science Foundation under Grant No.~DMR 1206553. Single crystalline X-ray diffractions by WX at LSU was supported by LSU startup funding.

\bibliography{endnote2}

\begin{thebibliography}{33}%
\makeatletter
\providecommand \@ifxundefined [1]{%
 \@ifx{#1\undefined}
}%
\providecommand \@ifnum [1]{%
 \ifnum #1\expandafter \@firstoftwo
 \else \expandafter \@secondoftwo
 \fi
}%
\providecommand \@ifx [1]{%
 \ifx #1\expandafter \@firstoftwo
 \else \expandafter \@secondoftwo
 \fi
}%
\providecommand \natexlab [1]{#1}%
\providecommand \enquote  [1]{``#1''}%
\providecommand \bibnamefont  [1]{#1}%
\providecommand \bibfnamefont [1]{#1}%
\providecommand \citenamefont [1]{#1}%
\providecommand \href@noop [0]{\@secondoftwo}%
\providecommand \href [0]{\begingroup \@sanitize@url \@href}%
\providecommand \@href[1]{\@@startlink{#1}\@@href}%
\providecommand \@@href[1]{\endgroup#1\@@endlink}%
\providecommand \@sanitize@url [0]{\catcode `\\12\catcode `\$12\catcode
  `\&12\catcode `\#12\catcode `\^12\catcode `\_12\catcode `\%12\relax}%
\providecommand \@@startlink[1]{}%
\providecommand \@@endlink[0]{}%
\providecommand \url  [0]{\begingroup\@sanitize@url \@url }%
\providecommand \@url [1]{\endgroup\@href {#1}{\urlprefix }}%
\providecommand \urlprefix  [0]{URL }%
\providecommand \Eprint [0]{\href }%
\providecommand \doibase [0]{http://dx.doi.org/}%
\providecommand \selectlanguage [0]{\@gobble}%
\providecommand \bibinfo  [0]{\@secondoftwo}%
\providecommand \bibfield  [0]{\@secondoftwo}%
\providecommand \translation [1]{[#1]}%
\providecommand \BibitemOpen [0]{}%
\providecommand \bibitemStop [0]{}%
\providecommand \bibitemNoStop [0]{.\EOS\space}%
\providecommand \EOS [0]{\spacefactor3000\relax}%
\providecommand \BibitemShut  [1]{\csname bibitem#1\endcsname}%
\let\auto@bib@innerbib\@empty
\bibitem [{\citenamefont {Jaccarino}\ \emph {et~al.}(1967)\citenamefont
  {Jaccarino}, \citenamefont {Wertheim}, \citenamefont {Wernick}, \citenamefont
  {Walker},\ and\ \citenamefont {Arajs}}]{jaccarino1967paramagnetic}%
  \BibitemOpen
  \bibfield  {author} {\bibinfo {author} {\bibfnamefont {V.}~\bibnamefont
  {Jaccarino}}, \bibinfo {author} {\bibfnamefont {G.~K.}\ \bibnamefont
  {Wertheim}}, \bibinfo {author} {\bibfnamefont {J.~H.}\ \bibnamefont
  {Wernick}}, \bibinfo {author} {\bibfnamefont {L.~R.}\ \bibnamefont {Walker}},
  \ and\ \bibinfo {author} {\bibfnamefont {S.}~\bibnamefont {Arajs}},\
  }\href@noop {} {\bibfield  {journal} {\bibinfo  {journal} {Physical Review}\
  }\textbf {\bibinfo {volume} {160}},\ \bibinfo {pages} {476} (\bibinfo {year}
  {1967})}\BibitemShut {NoStop}%
\bibitem [{\citenamefont {Z.~Schlesinger}\ and\ \citenamefont
  {Aeppli}(1993)}]{schlesinger1993h}%
  \BibitemOpen
  \bibfield  {author} {\bibinfo {author} {\bibfnamefont {H.-T. Z. M. B. M.
  J.~D.}\ \bibnamefont {Z.~Schlesinger}, \bibfnamefont {Z.~Fisk}}\ and\
  \bibinfo {author} {\bibfnamefont {G.}~\bibnamefont {Aeppli}},\ }\href@noop {}
  {\bibfield  {journal} {\bibinfo  {journal} {Phys. Rev. Lett}\ }\textbf
  {\bibinfo {volume} {71}},\ \bibinfo {pages} {1748} (\bibinfo {year}
  {1993})}\BibitemShut {NoStop}%
\bibitem [{\citenamefont {Sales}\ \emph {et~al.}(1994)\citenamefont {Sales},
  \citenamefont {Jones}, \citenamefont {Chakoumakos}, \citenamefont
  {Fernandez-Baca}, \citenamefont {Harmon}, \citenamefont {Sharp},\ and\
  \citenamefont {Volckmann}}]{sales1994magnetic}%
  \BibitemOpen
  \bibfield  {author} {\bibinfo {author} {\bibfnamefont {B.~C.}\ \bibnamefont
  {Sales}}, \bibinfo {author} {\bibfnamefont {E.~C.}\ \bibnamefont {Jones}},
  \bibinfo {author} {\bibfnamefont {B.~C.}\ \bibnamefont {Chakoumakos}},
  \bibinfo {author} {\bibfnamefont {J.~A.}\ \bibnamefont {Fernandez-Baca}},
  \bibinfo {author} {\bibfnamefont {H.~E.}\ \bibnamefont {Harmon}}, \bibinfo
  {author} {\bibfnamefont {J.~W.}\ \bibnamefont {Sharp}}, \ and\ \bibinfo
  {author} {\bibfnamefont {E.~H.}\ \bibnamefont {Volckmann}},\ }\href@noop {}
  {\bibfield  {journal} {\bibinfo  {journal} {Physical Review B}\ }\textbf
  {\bibinfo {volume} {50}},\ \bibinfo {pages} {8207} (\bibinfo {year}
  {1994})}\BibitemShut {NoStop}%
\bibitem [{\citenamefont {Paschen}\ \emph {et~al.}(1997)\citenamefont
  {Paschen}, \citenamefont {Felder}, \citenamefont {Chernikov}, \citenamefont
  {Degiorgi}, \citenamefont {Schwer}, \citenamefont {Ott}, \citenamefont
  {Young}, \citenamefont {Sarrao},\ and\ \citenamefont
  {Fisk}}]{paschen1997low}%
  \BibitemOpen
  \bibfield  {author} {\bibinfo {author} {\bibfnamefont {S.}~\bibnamefont
  {Paschen}}, \bibinfo {author} {\bibfnamefont {E.}~\bibnamefont {Felder}},
  \bibinfo {author} {\bibfnamefont {M.~A.}\ \bibnamefont {Chernikov}}, \bibinfo
  {author} {\bibfnamefont {L.}~\bibnamefont {Degiorgi}}, \bibinfo {author}
  {\bibfnamefont {H.}~\bibnamefont {Schwer}}, \bibinfo {author} {\bibfnamefont
  {H.~R.}\ \bibnamefont {Ott}}, \bibinfo {author} {\bibfnamefont {D.~P.}\
  \bibnamefont {Young}}, \bibinfo {author} {\bibfnamefont {J.~L.}\ \bibnamefont
  {Sarrao}}, \ and\ \bibinfo {author} {\bibfnamefont {Z.}~\bibnamefont
  {Fisk}},\ }\href@noop {} {\bibfield  {journal} {\bibinfo  {journal} {Physical
  Review B}\ }\textbf {\bibinfo {volume} {56}},\ \bibinfo {pages} {12916}
  (\bibinfo {year} {1997})}\BibitemShut {NoStop}%
\bibitem [{\citenamefont {Takahashi}\ and\ \citenamefont
  {Moriya}(1979)}]{takahashi1979theory}%
  \BibitemOpen
  \bibfield  {author} {\bibinfo {author} {\bibfnamefont {Y.}~\bibnamefont
  {Takahashi}}\ and\ \bibinfo {author} {\bibfnamefont {T.}~\bibnamefont
  {Moriya}},\ }\href@noop {} {\bibfield  {journal} {\bibinfo  {journal}
  {Journal of the Physical Society of Japan}\ }\textbf {\bibinfo {volume}
  {46}},\ \bibinfo {pages} {1451} (\bibinfo {year} {1979})}\BibitemShut
  {NoStop}%
\bibitem [{\citenamefont {Aeppli}\ and\ \citenamefont
  {Fisk}(1992)}]{aeppli1992comments}%
  \BibitemOpen
  \bibfield  {author} {\bibinfo {author} {\bibfnamefont {G.}~\bibnamefont
  {Aeppli}}\ and\ \bibinfo {author} {\bibfnamefont {Z.}~\bibnamefont {Fisk}},\
  }\href@noop {} {\bibfield  {journal} {\bibinfo  {journal} {Matter Phys}\
  }\textbf {\bibinfo {volume} {16}},\ \bibinfo {pages} {1192} (\bibinfo {year}
  {1992})}\BibitemShut {NoStop}%
\bibitem [{\citenamefont {Fisk}\ \emph {et~al.}(1995)\citenamefont {Fisk},
  \citenamefont {Sarrao}, \citenamefont {Thompson}, \citenamefont {Mandrus},
  \citenamefont {Hundley}, \citenamefont {Miglori}, \citenamefont {Bucher},
  \citenamefont {Schlesinger}, \citenamefont {Aeppli}, \citenamefont {Bucher},
  \citenamefont {DiTusa}, \citenamefont {Oglesby}, \citenamefont {Ott},
  \citenamefont {Canfield},\ and\ \citenamefont {Brown}}]{fisk1995kondo}%
  \BibitemOpen
  \bibfield  {author} {\bibinfo {author} {\bibfnamefont {Z.}~\bibnamefont
  {Fisk}}, \bibinfo {author} {\bibfnamefont {J.~L.}\ \bibnamefont {Sarrao}},
  \bibinfo {author} {\bibfnamefont {J.~D.}\ \bibnamefont {Thompson}}, \bibinfo
  {author} {\bibfnamefont {D.}~\bibnamefont {Mandrus}}, \bibinfo {author}
  {\bibfnamefont {M.~F.}\ \bibnamefont {Hundley}}, \bibinfo {author}
  {\bibfnamefont {A.}~\bibnamefont {Miglori}}, \bibinfo {author} {\bibfnamefont
  {B.}~\bibnamefont {Bucher}}, \bibinfo {author} {\bibfnamefont
  {Z.}~\bibnamefont {Schlesinger}}, \bibinfo {author} {\bibfnamefont
  {G.}~\bibnamefont {Aeppli}}, \bibinfo {author} {\bibfnamefont
  {E.}~\bibnamefont {Bucher}}, \bibinfo {author} {\bibfnamefont {J.~F.}\
  \bibnamefont {DiTusa}}, \bibinfo {author} {\bibfnamefont {C.~S.}\
  \bibnamefont {Oglesby}}, \bibinfo {author} {\bibfnamefont {H.~R.}\
  \bibnamefont {Ott}}, \bibinfo {author} {\bibfnamefont {P.}~\bibnamefont
  {Canfield}}, \ and\ \bibinfo {author} {\bibfnamefont {S.~E.}\ \bibnamefont
  {Brown}},\ }\href@noop {} {\bibfield  {journal} {\bibinfo  {journal} {Physica
  B: Condensed Matter}\ }\textbf {\bibinfo {volume} {206}},\ \bibinfo {pages}
  {798} (\bibinfo {year} {1995})}\BibitemShut {NoStop}%
\bibitem [{\citenamefont {Varma}(1994)}]{varma1994aspects}%
  \BibitemOpen
  \bibfield  {author} {\bibinfo {author} {\bibfnamefont {C.~M.}\ \bibnamefont
  {Varma}},\ }\href@noop {} {\bibfield  {journal} {\bibinfo  {journal}
  {Physical Review B}\ }\textbf {\bibinfo {volume} {50}},\ \bibinfo {pages}
  {9952} (\bibinfo {year} {1994})}\BibitemShut {NoStop}%
\bibitem [{\citenamefont {Fu}\ and\ \citenamefont
  {Doniach}(1995)}]{fu1995model}%
  \BibitemOpen
  \bibfield  {author} {\bibinfo {author} {\bibfnamefont {C.}~\bibnamefont
  {Fu}}\ and\ \bibinfo {author} {\bibfnamefont {S.}~\bibnamefont {Doniach}},\
  }\href@noop {} {\bibfield  {journal} {\bibinfo  {journal} {Physical Review
  B}\ }\textbf {\bibinfo {volume} {51}},\ \bibinfo {pages} {17439} (\bibinfo
  {year} {1995})}\BibitemShut {NoStop}%
\bibitem [{\citenamefont {Anisimov}\ \emph {et~al.}(1996)\citenamefont
  {Anisimov}, \citenamefont {Ezhov}, \citenamefont {Elfimov}, \citenamefont
  {Solovyev},\ and\ \citenamefont {Rice}}]{anisimov1996singlet}%
  \BibitemOpen
  \bibfield  {author} {\bibinfo {author} {\bibfnamefont {V.~I.}\ \bibnamefont
  {Anisimov}}, \bibinfo {author} {\bibfnamefont {S.~Y.}\ \bibnamefont {Ezhov}},
  \bibinfo {author} {\bibfnamefont {I.~S.}\ \bibnamefont {Elfimov}}, \bibinfo
  {author} {\bibfnamefont {I.~V.}\ \bibnamefont {Solovyev}}, \ and\ \bibinfo
  {author} {\bibfnamefont {T.~M.}\ \bibnamefont {Rice}},\ }\href@noop {}
  {\bibfield  {journal} {\bibinfo  {journal} {Physical review letters}\
  }\textbf {\bibinfo {volume} {76}},\ \bibinfo {pages} {1735} (\bibinfo {year}
  {1996})}\BibitemShut {NoStop}%
\bibitem [{\citenamefont {Wertheim}\ \emph {et~al.}(1965)\citenamefont
  {Wertheim}, \citenamefont {Jaccarino}, \citenamefont {Wernick}, \citenamefont
  {Seitchik}, \citenamefont {Williams},\ and\ \citenamefont
  {Sherwood}}]{wertheim1965unusual}%
  \BibitemOpen
  \bibfield  {author} {\bibinfo {author} {\bibfnamefont {G.~K.}\ \bibnamefont
  {Wertheim}}, \bibinfo {author} {\bibfnamefont {V.}~\bibnamefont {Jaccarino}},
  \bibinfo {author} {\bibfnamefont {J.~H.}\ \bibnamefont {Wernick}}, \bibinfo
  {author} {\bibfnamefont {J.~A.}\ \bibnamefont {Seitchik}}, \bibinfo {author}
  {\bibfnamefont {H.~J.}\ \bibnamefont {Williams}}, \ and\ \bibinfo {author}
  {\bibfnamefont {R.~C.}\ \bibnamefont {Sherwood}},\ }\href@noop {} {\bibfield
  {journal} {\bibinfo  {journal} {Physics Letters}\ }\textbf {\bibinfo {volume}
  {18}},\ \bibinfo {pages} {89} (\bibinfo {year} {1965})}\BibitemShut {NoStop}%
\bibitem [{\citenamefont {Kohgi}\ and\ \citenamefont
  {Ishikawa}(1981)}]{kohgi1981neutron}%
  \BibitemOpen
  \bibfield  {author} {\bibinfo {author} {\bibfnamefont {M.}~\bibnamefont
  {Kohgi}}\ and\ \bibinfo {author} {\bibfnamefont {Y.}~\bibnamefont
  {Ishikawa}},\ }\href@noop {} {\bibfield  {journal} {\bibinfo  {journal}
  {Solid State Communications}\ }\textbf {\bibinfo {volume} {37}},\ \bibinfo
  {pages} {833} (\bibinfo {year} {1981})}\BibitemShut {NoStop}%
\bibitem [{\citenamefont {Jarlborg}(1995)}]{jarlborg1995low}%
  \BibitemOpen
  \bibfield  {author} {\bibinfo {author} {\bibfnamefont {T.}~\bibnamefont
  {Jarlborg}},\ }\href@noop {} {\bibfield  {journal} {\bibinfo  {journal}
  {Physical Review B}\ }\textbf {\bibinfo {volume} {51}},\ \bibinfo {pages}
  {11106} (\bibinfo {year} {1995})}\BibitemShut {NoStop}%
\bibitem [{\citenamefont {Sun}\ \emph {et~al.}(2014)\citenamefont {Sun},
  \citenamefont {Wei}, \citenamefont {Menzel},\ and\ \citenamefont
  {Steglich}}]{sun2014resonant}%
  \BibitemOpen
  \bibfield  {author} {\bibinfo {author} {\bibfnamefont {P.}~\bibnamefont
  {Sun}}, \bibinfo {author} {\bibfnamefont {B.}~\bibnamefont {Wei}}, \bibinfo
  {author} {\bibfnamefont {D.}~\bibnamefont {Menzel}}, \ and\ \bibinfo {author}
  {\bibfnamefont {F.}~\bibnamefont {Steglich}},\ }\href@noop {} {\bibfield
  {journal} {\bibinfo  {journal} {Physical Review B}\ }\textbf {\bibinfo
  {volume} {90}},\ \bibinfo {pages} {245146} (\bibinfo {year}
  {2014})}\BibitemShut {NoStop}%
\bibitem [{\citenamefont {Petrova}\ \emph {et~al.}(2010)\citenamefont
  {Petrova}, \citenamefont {Krasnorussky}, \citenamefont {Shikov},
  \citenamefont {Yuhasz}, \citenamefont {Lograsso}, \citenamefont {Lashley},\
  and\ \citenamefont {Stishov}}]{petrova2010elastic}%
  \BibitemOpen
  \bibfield  {author} {\bibinfo {author} {\bibfnamefont {A.~E.}\ \bibnamefont
  {Petrova}}, \bibinfo {author} {\bibfnamefont {V.~N.}\ \bibnamefont
  {Krasnorussky}}, \bibinfo {author} {\bibfnamefont {A.~A.}\ \bibnamefont
  {Shikov}}, \bibinfo {author} {\bibfnamefont {W.~M.}\ \bibnamefont {Yuhasz}},
  \bibinfo {author} {\bibfnamefont {T.~A.}\ \bibnamefont {Lograsso}}, \bibinfo
  {author} {\bibfnamefont {J.~C.}\ \bibnamefont {Lashley}}, \ and\ \bibinfo
  {author} {\bibfnamefont {S.~M.}\ \bibnamefont {Stishov}},\ }\href@noop {}
  {\bibfield  {journal} {\bibinfo  {journal} {Physical Review B}\ }\textbf
  {\bibinfo {volume} {82}},\ \bibinfo {pages} {155124} (\bibinfo {year}
  {2010})}\BibitemShut {NoStop}%
\bibitem [{\citenamefont {Mihalik}\ \emph {et~al.}(1996)\citenamefont
  {Mihalik}, \citenamefont {Timko}, \citenamefont {Samuely}, \citenamefont
  {Toma{\v{s}}ovi{\v{c}}ova-Hud{\'a}kova}, \citenamefont {Szab{\'o}},\ and\
  \citenamefont {Menovsky}}]{mihalik1996magnetic}%
  \BibitemOpen
  \bibfield  {author} {\bibinfo {author} {\bibfnamefont {M.}~\bibnamefont
  {Mihalik}}, \bibinfo {author} {\bibfnamefont {M.}~\bibnamefont {Timko}},
  \bibinfo {author} {\bibfnamefont {P.}~\bibnamefont {Samuely}}, \bibinfo
  {author} {\bibfnamefont {N.}~\bibnamefont
  {Toma{\v{s}}ovi{\v{c}}ova-Hud{\'a}kova}}, \bibinfo {author} {\bibfnamefont
  {P.}~\bibnamefont {Szab{\'o}}}, \ and\ \bibinfo {author} {\bibfnamefont
  {A.~A.}\ \bibnamefont {Menovsky}},\ }\href@noop {} {\bibfield  {journal}
  {\bibinfo  {journal} {Journal of magnetism and magnetic materials}\ }\textbf
  {\bibinfo {volume} {157}},\ \bibinfo {pages} {637} (\bibinfo {year}
  {1996})}\BibitemShut {NoStop}%
\bibitem [{\citenamefont {Bocellip}\ \emph {et~al.}(1995)\citenamefont
  {Bocellip}, \citenamefont {Marabelli}, \citenamefont {Spolenak},\ and\
  \citenamefont {Bauer}}]{bocellip1995evolution}%
  \BibitemOpen
  \bibfield  {author} {\bibinfo {author} {\bibfnamefont {S.}~\bibnamefont
  {Bocellip}}, \bibinfo {author} {\bibfnamefont {F.}~\bibnamefont {Marabelli}},
  \bibinfo {author} {\bibfnamefont {R.}~\bibnamefont {Spolenak}}, \ and\
  \bibinfo {author} {\bibfnamefont {E.}~\bibnamefont {Bauer}},\ }\href@noop {}
  {\bibfield  {journal} {\bibinfo  {journal} {MRS Online Proceedings Library
  Archive}\ }\textbf {\bibinfo {volume} {402}} (\bibinfo {year}
  {1995})}\BibitemShut {NoStop}%
\bibitem [{\citenamefont {Schlesinger}\ \emph {et~al.}(1993)\citenamefont
  {Schlesinger}, \citenamefont {Fisk}, \citenamefont {Zhang}, \citenamefont
  {Maple}, \citenamefont {DiTusa},\ and\ \citenamefont
  {Aeppli}}]{schlesinger1993unconventional}%
  \BibitemOpen
  \bibfield  {author} {\bibinfo {author} {\bibfnamefont {Z.~.}\ \bibnamefont
  {Schlesinger}}, \bibinfo {author} {\bibfnamefont {Z.}~\bibnamefont {Fisk}},
  \bibinfo {author} {\bibfnamefont {H.-T.}\ \bibnamefont {Zhang}}, \bibinfo
  {author} {\bibfnamefont {M.}~\bibnamefont {Maple}}, \bibinfo {author}
  {\bibfnamefont {J.}~\bibnamefont {DiTusa}}, \ and\ \bibinfo {author}
  {\bibfnamefont {G.}~\bibnamefont {Aeppli}},\ }\href@noop {} {\bibfield
  {journal} {\bibinfo  {journal} {Physical review letters}\ }\textbf {\bibinfo
  {volume} {71}},\ \bibinfo {pages} {1748} (\bibinfo {year}
  {1993})}\BibitemShut {NoStop}%
\bibitem [{\citenamefont {Hunt}\ \emph {et~al.}(1994)\citenamefont {Hunt},
  \citenamefont {Chernikov}, \citenamefont {Felder}, \citenamefont {Ott},
  \citenamefont {Fisk},\ and\ \citenamefont {Canfield}}]{hunt1994low}%
  \BibitemOpen
  \bibfield  {author} {\bibinfo {author} {\bibfnamefont {M.~B.}\ \bibnamefont
  {Hunt}}, \bibinfo {author} {\bibfnamefont {M.~A.}\ \bibnamefont {Chernikov}},
  \bibinfo {author} {\bibfnamefont {E.}~\bibnamefont {Felder}}, \bibinfo
  {author} {\bibfnamefont {H.~R.}\ \bibnamefont {Ott}}, \bibinfo {author}
  {\bibfnamefont {Z.}~\bibnamefont {Fisk}}, \ and\ \bibinfo {author}
  {\bibfnamefont {P.}~\bibnamefont {Canfield}},\ }\href@noop {} {\bibfield
  {journal} {\bibinfo  {journal} {Physical Review B}\ }\textbf {\bibinfo
  {volume} {50}},\ \bibinfo {pages} {14933} (\bibinfo {year}
  {1994})}\BibitemShut {NoStop}%
\bibitem [{\citenamefont {Brian}\ \emph {et~al.}(2011)\citenamefont {Brian},
  \citenamefont {Olivier}, \citenamefont {Michael},\ and\ \citenamefont
  {May}}]{sales2011thermoelectric}%
  \BibitemOpen
  \bibfield  {author} {\bibinfo {author} {\bibfnamefont {C.~S.}\ \bibnamefont
  {Brian}}, \bibinfo {author} {\bibfnamefont {D.}~\bibnamefont {Olivier}},
  \bibinfo {author} {\bibfnamefont {A.~M.}\ \bibnamefont {Michael}}, \ and\
  \bibinfo {author} {\bibfnamefont {A.~F.}\ \bibnamefont {May}},\ }\href@noop
  {} {\bibfield  {journal} {\bibinfo  {journal} {Physical review B}\ }\textbf
  {\bibinfo {volume} {83}} (\bibinfo {year} {2011})}\BibitemShut {NoStop}%
\bibitem [{\citenamefont {Dietl}(2010)}]{dietl2010a}%
  \BibitemOpen
  \bibfield  {author} {\bibinfo {author} {\bibfnamefont {T.}~\bibnamefont
  {Dietl}},\ }\href@noop {} {\bibfield  {journal} {\bibinfo  {journal} {Nature
  Materials}\ }\textbf {\bibinfo {volume} {9}},\ \bibinfo {pages} {965}
  (\bibinfo {year} {2010})}\BibitemShut {NoStop}%
\bibitem [{\citenamefont {Fang}\ \emph
  {et~al.}(2015{\natexlab{a}})\citenamefont {Fang}, \citenamefont {Wolowiec},
  \citenamefont {Yazici},\ and\ \citenamefont {Maple}}]{fang2015chemical}%
  \BibitemOpen
  \bibfield  {author} {\bibinfo {author} {\bibfnamefont {Y.}~\bibnamefont
  {Fang}}, \bibinfo {author} {\bibfnamefont {C.~T.}\ \bibnamefont {Wolowiec}},
  \bibinfo {author} {\bibfnamefont {D.}~\bibnamefont {Yazici}}, \ and\ \bibinfo
  {author} {\bibfnamefont {M.~B.}\ \bibnamefont {Maple}},\ }\href@noop {}
  {\bibfield  {journal} {\bibinfo  {journal} {Novel Superconducting Materials}\
  }\textbf {\bibinfo {volume} {1}},\ \bibinfo {pages} {79} (\bibinfo {year}
  {2015}{\natexlab{a}})}\BibitemShut {NoStop}%
\bibitem [{\citenamefont {Sheldrick}(2008)}]{sheldrick2008short}%
  \BibitemOpen
  \bibfield  {author} {\bibinfo {author} {\bibfnamefont {G.~M.}\ \bibnamefont
  {Sheldrick}},\ }\href@noop {} {\bibfield  {journal} {\bibinfo  {journal}
  {Acta Crystallographica Section A: Foundations of Crystallography}\ }\textbf
  {\bibinfo {volume} {64}},\ \bibinfo {pages} {112} (\bibinfo {year}
  {2008})}\BibitemShut {NoStop}%
\bibitem [{\citenamefont {Fang}\ \emph
  {et~al.}(2015{\natexlab{b}})\citenamefont {Fang}, \citenamefont {Yazici},
  \citenamefont {White},\ and\ \citenamefont {Maple}}]{fang2015enhancement}%
  \BibitemOpen
  \bibfield  {author} {\bibinfo {author} {\bibfnamefont {Y.}~\bibnamefont
  {Fang}}, \bibinfo {author} {\bibfnamefont {D.}~\bibnamefont {Yazici}},
  \bibinfo {author} {\bibfnamefont {B.~D.}\ \bibnamefont {White}}, \ and\
  \bibinfo {author} {\bibfnamefont {M.~B.}\ \bibnamefont {Maple}},\ }\href@noop
  {} {\bibfield  {journal} {\bibinfo  {journal} {Physical review B}\ }\textbf
  {\bibinfo {volume} {91}} (\bibinfo {year} {2015}{\natexlab{b}})}\BibitemShut
  {NoStop}%
\bibitem [{\citenamefont {Samuely}\ \emph {et~al.}(1996)\citenamefont
  {Samuely}, \citenamefont {Szab{\'o}}, \citenamefont {Mihalik}, \citenamefont
  {Hud{\'a}kov{\'a}},\ and\ \citenamefont {Menovsky}}]{samuely1996gap}%
  \BibitemOpen
  \bibfield  {author} {\bibinfo {author} {\bibfnamefont {P.}~\bibnamefont
  {Samuely}}, \bibinfo {author} {\bibfnamefont {P.}~\bibnamefont {Szab{\'o}}},
  \bibinfo {author} {\bibfnamefont {M.}~\bibnamefont {Mihalik}}, \bibinfo
  {author} {\bibfnamefont {N.}~\bibnamefont {Hud{\'a}kov{\'a}}}, \ and\
  \bibinfo {author} {\bibfnamefont {A.~A.}\ \bibnamefont {Menovsky}},\
  }\href@noop {} {\bibfield  {journal} {\bibinfo  {journal} {Physica B:
  Condensed Matter}\ }\textbf {\bibinfo {volume} {218}},\ \bibinfo {pages}
  {185} (\bibinfo {year} {1996})}\BibitemShut {NoStop}%
\bibitem [{\citenamefont {Wolfe}\ \emph {et~al.}(1965)\citenamefont {Wolfe},
  \citenamefont {Wernick},\ and\ \citenamefont
  {Haszko}}]{wolfe1965thermoelectric}%
  \BibitemOpen
  \bibfield  {author} {\bibinfo {author} {\bibfnamefont {R.}~\bibnamefont
  {Wolfe}}, \bibinfo {author} {\bibfnamefont {J.~H.}\ \bibnamefont {Wernick}},
  \ and\ \bibinfo {author} {\bibfnamefont {S.~E.}\ \bibnamefont {Haszko}},\
  }\href@noop {} {\bibfield  {journal} {\bibinfo  {journal} {Physics Letters}\
  }\textbf {\bibinfo {volume} {19}},\ \bibinfo {pages} {449} (\bibinfo {year}
  {1965})}\BibitemShut {NoStop}%
\bibitem [{\citenamefont {Buschinger}\ \emph {et~al.}(1997)\citenamefont
  {Buschinger}, \citenamefont {Geibel}, \citenamefont {Steglich}, \citenamefont
  {Mandrus}, \citenamefont {Young}, \citenamefont {Sarrao},\ and\ \citenamefont
  {Fisk}}]{buschinger1997transport}%
  \BibitemOpen
  \bibfield  {author} {\bibinfo {author} {\bibfnamefont {B.}~\bibnamefont
  {Buschinger}}, \bibinfo {author} {\bibfnamefont {C.}~\bibnamefont {Geibel}},
  \bibinfo {author} {\bibfnamefont {F.}~\bibnamefont {Steglich}}, \bibinfo
  {author} {\bibfnamefont {D.}~\bibnamefont {Mandrus}}, \bibinfo {author}
  {\bibfnamefont {D.}~\bibnamefont {Young}}, \bibinfo {author} {\bibfnamefont
  {J.~L.}\ \bibnamefont {Sarrao}}, \ and\ \bibinfo {author} {\bibfnamefont
  {Z.}~\bibnamefont {Fisk}},\ }\href@noop {} {\bibfield  {journal} {\bibinfo
  {journal} {Physica B: Condensed Matter}\ }\textbf {\bibinfo {volume} {230}},\
  \bibinfo {pages} {784} (\bibinfo {year} {1997})}\BibitemShut {NoStop}%
\bibitem [{\citenamefont {Ishizaka}\ \emph {et~al.}(2005)\citenamefont
  {Ishizaka}, \citenamefont {Kiss}, \citenamefont {Shimojima}, \citenamefont
  {Yokoya}, \citenamefont {Togashi}, \citenamefont {Watanabe}, \citenamefont
  {Zhang}, \citenamefont {Chen}, \citenamefont {Onose}, \citenamefont
  {Tokura},\ and\ \citenamefont {Shin}}]{ishizaka2005ultraviolet}%
  \BibitemOpen
  \bibfield  {author} {\bibinfo {author} {\bibfnamefont {K.}~\bibnamefont
  {Ishizaka}}, \bibinfo {author} {\bibfnamefont {T.}~\bibnamefont {Kiss}},
  \bibinfo {author} {\bibfnamefont {T.}~\bibnamefont {Shimojima}}, \bibinfo
  {author} {\bibfnamefont {T.}~\bibnamefont {Yokoya}}, \bibinfo {author}
  {\bibfnamefont {T.}~\bibnamefont {Togashi}}, \bibinfo {author} {\bibfnamefont
  {S.}~\bibnamefont {Watanabe}}, \bibinfo {author} {\bibfnamefont {C.~Q.}\
  \bibnamefont {Zhang}}, \bibinfo {author} {\bibfnamefont {C.~T.}\ \bibnamefont
  {Chen}}, \bibinfo {author} {\bibfnamefont {Y.}~\bibnamefont {Onose}},
  \bibinfo {author} {\bibfnamefont {Y.}~\bibnamefont {Tokura}}, \ and\ \bibinfo
  {author} {\bibfnamefont {S.}~\bibnamefont {Shin}},\ }\href@noop {} {\bibfield
   {journal} {\bibinfo  {journal} {Physical Review B}\ }\textbf {\bibinfo
  {volume} {72}},\ \bibinfo {pages} {233202} (\bibinfo {year}
  {2005})}\BibitemShut {NoStop}%
\bibitem [{\citenamefont {Arita}\ \emph {et~al.}(2008)\citenamefont {Arita},
  \citenamefont {Shimada}, \citenamefont {Takeda}, \citenamefont {Nakatake},
  \citenamefont {Namatame}, \citenamefont {Taniguchi}, \citenamefont {Negishi},
  \citenamefont {Oguchi}, \citenamefont {Saitoh}, \citenamefont {Fujimori},\
  and\ \citenamefont {Kanomata}}]{arita2008angle}%
  \BibitemOpen
  \bibfield  {author} {\bibinfo {author} {\bibfnamefont {M.}~\bibnamefont
  {Arita}}, \bibinfo {author} {\bibfnamefont {K.}~\bibnamefont {Shimada}},
  \bibinfo {author} {\bibfnamefont {Y.}~\bibnamefont {Takeda}}, \bibinfo
  {author} {\bibfnamefont {M.}~\bibnamefont {Nakatake}}, \bibinfo {author}
  {\bibfnamefont {H.}~\bibnamefont {Namatame}}, \bibinfo {author}
  {\bibfnamefont {M.}~\bibnamefont {Taniguchi}}, \bibinfo {author}
  {\bibfnamefont {H.}~\bibnamefont {Negishi}}, \bibinfo {author} {\bibfnamefont
  {T.}~\bibnamefont {Oguchi}}, \bibinfo {author} {\bibfnamefont
  {T.}~\bibnamefont {Saitoh}}, \bibinfo {author} {\bibfnamefont
  {A.}~\bibnamefont {Fujimori}}, \ and\ \bibinfo {author} {\bibfnamefont
  {T.}~\bibnamefont {Kanomata}},\ }\href@noop {} {\bibfield  {journal}
  {\bibinfo  {journal} {Physical Review B}\ }\textbf {\bibinfo {volume} {77}},\
  \bibinfo {pages} {205117} (\bibinfo {year} {2008})}\BibitemShut {NoStop}%
\bibitem [{\citenamefont {Klein}\ \emph {et~al.}(2008)\citenamefont {Klein},
  \citenamefont {Zur}, \citenamefont {Menzel}, \citenamefont {Schoenes},
  \citenamefont {Doll}, \citenamefont {R{\"o}der},\ and\ \citenamefont
  {Reinert}}]{klein2008evidence}%
  \BibitemOpen
  \bibfield  {author} {\bibinfo {author} {\bibfnamefont {M.}~\bibnamefont
  {Klein}}, \bibinfo {author} {\bibfnamefont {D.}~\bibnamefont {Zur}}, \bibinfo
  {author} {\bibfnamefont {D.}~\bibnamefont {Menzel}}, \bibinfo {author}
  {\bibfnamefont {J.}~\bibnamefont {Schoenes}}, \bibinfo {author}
  {\bibfnamefont {K.}~\bibnamefont {Doll}}, \bibinfo {author} {\bibfnamefont
  {J.}~\bibnamefont {R{\"o}der}}, \ and\ \bibinfo {author} {\bibfnamefont
  {F.}~\bibnamefont {Reinert}},\ }\href@noop {} {\bibfield  {journal} {\bibinfo
   {journal} {Physical review letters}\ }\textbf {\bibinfo {volume} {101}},\
  \bibinfo {pages} {046406} (\bibinfo {year} {2008})}\BibitemShut {NoStop}%
\bibitem [{\citenamefont {Takahashi}\ \emph {et~al.}(2000)\citenamefont
  {Takahashi}, \citenamefont {Kanomata}, \citenamefont {Note},\ and\
  \citenamefont {Nakagawa}}]{takahashi2000specific}%
  \BibitemOpen
  \bibfield  {author} {\bibinfo {author} {\bibfnamefont {Y.}~\bibnamefont
  {Takahashi}}, \bibinfo {author} {\bibfnamefont {T.}~\bibnamefont {Kanomata}},
  \bibinfo {author} {\bibfnamefont {R.}~\bibnamefont {Note}}, \ and\ \bibinfo
  {author} {\bibfnamefont {T.}~\bibnamefont {Nakagawa}},\ }\href@noop {}
  {\bibfield  {journal} {\bibinfo  {journal} {Journal of the Physical Society
  of Japan}\ }\textbf {\bibinfo {volume} {69}},\ \bibinfo {pages} {4018}
  (\bibinfo {year} {2000})}\BibitemShut {NoStop}%
\bibitem [{\citenamefont {Marklund}\ \emph {et~al.}(1974)\citenamefont
  {Marklund}, \citenamefont {Larsson}, \citenamefont {Bystr{\"o}m},\ and\
  \citenamefont {Lindqvist}}]{marklund1974specific}%
  \BibitemOpen
  \bibfield  {author} {\bibinfo {author} {\bibfnamefont {K.}~\bibnamefont
  {Marklund}}, \bibinfo {author} {\bibfnamefont {M.}~\bibnamefont {Larsson}},
  \bibinfo {author} {\bibfnamefont {S.}~\bibnamefont {Bystr{\"o}m}}, \ and\
  \bibinfo {author} {\bibfnamefont {T.}~\bibnamefont {Lindqvist}},\ }\href@noop
  {} {\bibfield  {journal} {\bibinfo  {journal} {Physica Scripta}\ }\textbf
  {\bibinfo {volume} {9}},\ \bibinfo {pages} {47} (\bibinfo {year}
  {1974})}\BibitemShut {NoStop}%
\bibitem [{\citenamefont {Koyama}\ \emph {et~al.}(1999)\citenamefont {Koyama},
  \citenamefont {Goto}, \citenamefont {Kanomata},\ and\ \citenamefont
  {Note}}]{koyama1999precise}%
  \BibitemOpen
  \bibfield  {author} {\bibinfo {author} {\bibfnamefont {K.}~\bibnamefont
  {Koyama}}, \bibinfo {author} {\bibfnamefont {T.}~\bibnamefont {Goto}},
  \bibinfo {author} {\bibfnamefont {T.}~\bibnamefont {Kanomata}}, \ and\
  \bibinfo {author} {\bibfnamefont {R.}~\bibnamefont {Note}},\ }\href@noop {}
  {\bibfield  {journal} {\bibinfo  {journal} {Journal of the Physical Society
  of Japan}\ }\textbf {\bibinfo {volume} {68}},\ \bibinfo {pages} {1693}
  (\bibinfo {year} {1999})}\BibitemShut {NoStop}%
\end{thebibliography}%
\clearpage
\end{document}